\documentclass[aps,pra,twocolumn,showpacs]{revtex4-1}

\usepackage{amsmath,amssymb}
\usepackage{graphicx}

\usepackage{color}

\graphicspath{ {.} {./pics/} }

\newcommand{\rmi}{\mathrm{i}}

\newcommand{\rme}{\mathrm{e}}

\begin{document}

\title{Multipartite entangled light from driven microcavities}

\author{D. Pagel}
%\email{pagel@physik.uni-greifswald.de}
\affiliation{Institut f\"ur Physik, Ernst-Moritz-Arndt-Universit\"at Greifswald, 17487 Greifswald, Germany}
\author{J. Sperling}
%\email{jan.sperling2@uni-rostock.de}
\affiliation{Institut f\"ur Physik, Universit\"at Rostock, 18051 Rostock, Germany}
\author{H. Fehske}
\email{fehske@physik.uni-greifswald.de}
\affiliation{Institut f\"ur Physik, Ernst-Moritz-Arndt-Universit\"at Greifswald, 17487 Greifswald, Germany}
\author{W. Vogel}
%\email{werner.vogel@uni-rostock.de}
\affiliation{Institut f\"ur Physik, Universit\"at Rostock, 18051 Rostock, Germany}

%_______________________________________________________________________________
%===============================================================================
\begin{abstract}
The generation and the characterization of multipartite entangled light is an important and challenging task in quantum optics.
In this paper the entanglement properties of the light emitted from a planar semiconductor microcavity are studied.
The intracavity scattering dynamics leads to the emission of light that is described by a fourpartite $W$ state.
Its multipartite correlations are identified by using the method of entanglement witnesses.
Entanglement conditions are derived, which are based on a general witness constructed from $W$ states.
The results can be used to detect entanglement of light that propagates through lossy and even turbulent media.
\end{abstract}


\maketitle

%_______________________________________________________________________________
%===============================================================================
\section{Introduction}

The phenomenon of quantum entanglement relies on the superposition principle of quantum physics.
Since the pioneering works~\cite{EPR35,Sch35} this effect has been regarded as one fundamental discrepancy between the quantum and classical domains of nature.
Nowadays, entanglement is considered to be a key resource for quantum information technologies, cf. e.g.~\cite{NC10,HHHH09,GT09}.

Quantum entanglement is defined as a kind of correlation between subsystems, which cannot be interpreted in terms of classical joint probabilities~\cite{Wer89}.
Especially in the multipartite scenario, these non-classical correlations exist in various forms, for an introduction see e.g.~\cite{HHHH09,GT09}.
The most elementary examples of non-equivalent forms are given by GHZ states~\cite{GHZ89} and $W$ states~\cite{DVC00}.
Among many possible applications of entanglement, the best studied ones are quantum key distribution~\cite{Ekert91}, quantum dense coding~\cite{BW92}, and quantum teleportation~\cite{BBCJPW93}.

Typically one can detect entanglement using so-called entanglement witnesses~\cite{HHH96,HHH01}.
These observables are non-negative for separable states, but exhibit negativities for entangled ones.
To quantify the amount of entanglement within a system~\cite{VPRK97,VP98,Vidal00,Bra05,BV06} one has to find a proper entanglement measure, which can be constructed from entanglement witnesses.
For bipartite systems the solution of such an optimization procedure is given~\cite{Bra05,SV11b}.
In the multipartite case, the problem of finding an optimal entanglement measure is still unsolved.
The construction of multipartite witnesses has been resolved only recently~\cite{SV13}.

A system, where the identification of multipartite entangled light becomes important is a two-dimensional semiconductor microcavity~\cite{Ciu04,Lang04,CBC05,AB12,SBSSWR00}.
Here, an optical driving with a laser field at a frequency near the fundamental band gap of the semiconductor can coherently create excitons, i.e., bound states of electrons and holes.
In the low density limit, excitons can be described as an ideal gas of bosons.
For high densities one has to account for the fermionic nature of the exciton constituents, leading to effective exciton-exciton interactions~\cite{TY99,CSQ01,OS98,Usui60,RCSPQS00}.
Within the microcavity, the strong coupling of cavity photons with semiconductor excitons leads to an anticrossing of the energy dispersions of the mixed exciton photon modes---so-called polaritons~\cite{WNIA92,HWSOPI94}.
Polariton-polariton interactions arise from the Coulomb interaction within their electronic parts~\cite{TY99,OS98}.
Due to this interaction, pumped polaritons can scatter into pairs of signal and idler polaritons, if energy and momentum are conserved.
It has been shown, that the signal and idler polaritons can be in an entangled state~\cite{Ciu04,SDSSL05,PDSSS09,EVWP13,PFSV12}.

While the generation of multipartite entanglement in planar microcavities is based on the strong coupling between the intracavity field and the semiconductor excitations, alternative generation schemes have been proposed in the literature.
Realizations involving linear optics such as beam splitters rely on parametric light sources, e.g.~squeezed light~\cite{DBCK10}.
Examples for setups using nonlinearities are concurrent interactions in second-order nonlinear media~\cite{PFJPX04}, interlinked interactions in $\chi^{(2)}$ media~\cite{FPBAPA04}, and down-conversion in parametric media~\cite{VLB00}.

In the present paper we demonstrate that multipartite entanglement can be created and identified in driven microcavities.
In particular, we consider the emitted light from a planar semiconductor microcavity that is driven by four pumps.
This leads to the generation of photons in a four-partite $W$ state.
The detection of their multipartite correlations is based on entanglement witnesses.
This method requires the solution of the so-called multipartite separability eigenvalue equations~\cite{SV13}, and we provide the full solution for a class of witnesses that is based on a generalized pure $W$ state.
This allows us to study the loss of entanglement of the emitted light when it propagates through lossy media.

We proceed as follows.
In Sec.~\ref{Sec:EntGen} we briefly recapitulate the bosonic description of planar microcavities and present the pump geometry that leads to the generation of polaritons in a $W$ state.
Their multipartite entanglement is verified in Sec.~\ref{Sec:IdentEnt}.
In Sec.~\ref{Sec:EntProp} we study the propagation of the emitted light through the atmosphere and its impact on the entanglement properties of the photons.
We use the solution to the separability eigenvalue equations for a generalized pure $W$-state witness, obtained in Sec.~\ref{Sec:WWitness}.
Section~\ref{Sec:Concl} presents our conclusions.

%_______________________________________________________________________________
%===============================================================================
\section{Setup for the generation of entangled light}\label{Sec:EntGen}
In this section, we give a short review of the description of planar microcavities in terms of bosonic polaritons~\cite{Ciu04,TY99,CSQ01}.
This description can easily be used to investigate polariton parametric scattering in momentum space~\cite{Ciu04,PFSV12}, and we propose a scenario that leads to the generation of polaritons in multipartite entangled states.

An alternative approach for the description of polariton scattering is based on equations of motion for the exciton and photon operators and is called dynamics controlled truncation formalism~\cite{AS94,SG96,PDSSRG08b}.
It was recently extended to double and triple cavities~\cite{EVWP13}.

\begin{figure*}
 \includegraphics{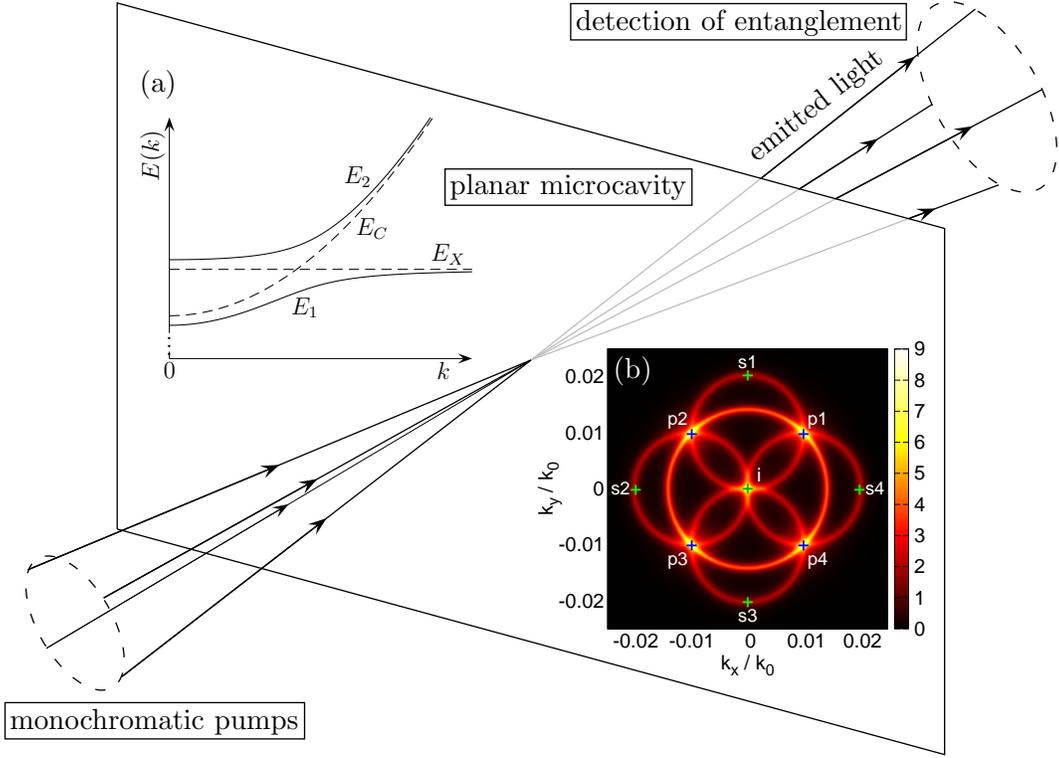}
 \caption{\label{Fig:setup}
  (Color online) Sketch of the considered physical processes.
  The pumping of the planar microcavity leads to the emission of light whose multipartite entanglement is detected.
  Inset (a) shows the energy dispersion relations of the excitons, cavity photons and polaritons.
  Inset (b) depicts the phase-matching function for $E_{\text{C}}(0) = E_{\text{X}} = 1.5\,$eV, $\Omega_{\text{R}} = 2\,$meV, and $\gamma = 10\,\mu$eV.
  The four pumps with $k_{\text{p}} = 0.01\,k_0$ are arranged on a cone.
 }
\end{figure*}

%_______________________________________________________________________________
%-------------------------------------------------------------------------------
\subsection{Bosonic description of planar microcavities}
Our staring point is the bosonic description of two-dimensional semiconductor microcavities in the basis of excitons and cavity photons.
The excitons are assumed to be dispersionless, i.e., $E_{\text{X}}(\mathbf{k}) = E_{\text{X}}$, whereas the photon energy $E_{\text{C}}$ grows linearly with the modulus of the three-dimensional wave vector.
Projected onto the two dimensions of the microcavity we obtain $E_{\text{C}}(k) = E_{\text{C}}(0) \sqrt{1 + (k / k_0)^2}$, where $k$ is the modulus of the in-plane wave vector $\mathbf{k}$ and $k_0 = E_{\text{C}}(0)$.
As a simplification, we work in units where $\hbar = c = 1$.

The interaction of excitons and cavity photons can be separated into a harmonic and an anharmonic contribution~\cite{CSPQS98,TY99}.
In the harmonic approximation, we can perform a Hopfield transformation~\cite{Hop58} to get new quasiparticles called polaritons.
The Hamiltonian of non-interacting polaritons then reads
\begin{equation}
 H_{\text{P}} = \sum_{j,\mathbf{k}} E_j(\mathbf{k}) p_{j \mathbf{k}}^\dagger p_{j \mathbf{k}}^{} \;,
\end{equation}
where $p_{j \mathbf{k}}^\dagger$ creates a polariton with in-plane wave vector $\mathbf{k}$ in the lower ($j = 1$) or upper ($j = 2$) branch with energy $E_j(\mathbf{k})$.
Figure~\ref{Fig:setup} inset~(a) schematically shows these functions (solid lines) together with the dispersions $E_{\text{C}}(k)$ and $E_{\text{X}}$ of the cavity photons and excitons (dashed lines).
For large values of $k$ the polariton modes are equal to the separated exciton and cavity photon modes.
For small $k$ the strong coupling between the excitons and the photons of the planar microcavity leads to an anticrossing of the polariton dispersions.

A polariton pair interaction arises from the anharmonic exciton photon coupling and from the Coulomb interaction within the electronic part of the excitons and is given by~\cite{RCSPQS00,Ciu04,TY99}
\begin{equation}\label{HPP}
 H_{\text{PP}} = \frac{1}{2} \sum_{\mathbf{k},\mathbf{k}',\mathbf{q}} \sum_{\substack{j_1,j_2\\j_3,j_4}} \frac{R_{\text{X}}^2}{A} V_{\mathbf{k},\mathbf{k}',\mathbf{q}}^{j_1 j_2 j_3 j_4} p_{j_1 \mathbf{k}+\mathbf{q}}^\dagger p_{j_2 \mathbf{k}'-\mathbf{q}}^\dagger p_{j_3 \mathbf{k}} p_{j_4 \mathbf{k}'} \;.
\end{equation}
In this equation $R_{\text{X}}$ is the exciton radius, $A$ is the sample surface, and $V_{\mathbf{k},\mathbf{k}',\mathbf{q}}^{j_1 j_2 j_3 j_4}$ is the effective branch-dependent potential,
\begin{eqnarray}
 \frac{V_{\mathbf{k},\mathbf{k}',\mathbf{q}}^{j_1 j_2 j_3 j_4}}{E_{\text{b}}} &=& 12 M_{1 j_1 \mathbf{k}+\mathbf{q}} M_{1 j_2 \mathbf{k}'-\mathbf{q}} M_{1 j_3 \mathbf{k}} M_{1 j_4 \mathbf{k}'} \nonumber\\*
 && -\frac{8 \pi}{7} p_{\text{s}} \big( M_{2 j_1 \mathbf{k}+\mathbf{q}} M_{1 j_2 \mathbf{k}'-\mathbf{q}} M_{1 j_3 \mathbf{k}} M_{1 j_4 \mathbf{k}'} \nonumber\\*
 && + M_{1 j_1 \mathbf{k}+\mathbf{q}} M_{1 j_2 \mathbf{k}'-\mathbf{q}} M_{1 j_3 \mathbf{k}} M_{2 j_4 \mathbf{k}'} \big) \;.
\end{eqnarray}
Here, $E_{\text{b}} = e^2 / (2 \epsilon R_{\text{X}})$ is the exciton binding energy with $\epsilon$ being the static dielectric constant of the crystal, and $p_{\text{s}} = 2 \Omega_{\text{R}} / E_{\text{b}}$ is the ratio of polariton splitting to binding energy.
The coefficients $M_{j j' \mathbf{k}}$ follow from the Hopfield transformation as $M_{1 1 \mathbf{k}} = M_{2 2 \mathbf{k}} = 1 / \sqrt{1 + \rho_{\mathbf{k}}^2}$ and $M_{1 2 \mathbf{k}} = -M_{2 1 \mathbf{k}} = \sqrt{1 - M_{1 1 \mathbf{k}}^2}$, where $\rho_{\mathbf{k}} = [E_2(\mathbf{k}) - E_{\text{C}}(\mathbf{k})] / \Omega_{\text{R}}$.

%_______________________________________________________________________________
%-------------------------------------------------------------------------------
\subsection{Polariton parametric scattering}
We consider an experimental setup that involves scattering processes within the lower polariton branch only.
We choose a four pump-scheme, where the wave vectors $\mathbf{k}_{\text{p}n}$ for $n = 1, \ldots, 4$ of the pumps have equal amplitudes.
The scattering of pumped polaritons into pairs of signal and idler is described by single-pump (signal at $\mathbf{k}$ and idler at $2\mathbf{k}_{\text{p}n} - \mathbf{k}$) and mixed-pump (signal at $\mathbf{k}$ and idler at $\mathbf{k}_{\text{p}n} + \mathbf{k}_{\text{p}m} - \mathbf{k}$ with $n \neq m$) parametric processes.
Within the setup under study (see Fig.~\ref{Fig:setup}), the incident angles of the four pumps shall be below the magic angle~\cite{Lang04,SBSSWR00}, such that single-pump parametric scattering is negligible.
In particular, we choose $\mathbf{k}_{\text{p}1} = (k_{\text{p}}, k_{\text{p}})$, $\mathbf{k}_{\text{p}2} = (-k_{\text{p}}, k_{\text{p}})$, $\mathbf{k}_{\text{p}3} = (-k_{\text{p}}, -k_{\text{p}})$, and $\mathbf{k}_{\text{p}4} = (k_{\text{p}}, -k_{\text{p}})$.

In Fig.~\ref{Fig:setup} inset~(b) we show the phase-matching function for this scenario.
This function is given by
\begin{eqnarray}
 \Phi(\mathbf{k}) &=& \sum_{n,m=1}^4 \gamma^2 \Big\{ \big[ E_1(k) + E_1(|\mathbf{k}_{\text{p}n} + \mathbf{k}_{\text{p}m} - \mathbf{k}|) \nonumber\\*
 && - 2 E_1(\sqrt{2} k_{\text{p}}) \big]^2 + \gamma^2 \Big\}^{-1} \;,
\end{eqnarray}
where $\gamma$ is the polariton broadening.
Mixed-pump scattering processes of oppositely arranged pumps ($|\mathbf{k}_{\text{p}n} + \mathbf{k}_{\text{p}m}| = 0$) contribute to the circle with radius $\sqrt{2} k_{\text{p}}$ in Fig.~\ref{Fig:setup} inset~(b).
The four mixed-pump processes of neighboring pumps ($|\mathbf{k}_{\text{p}n} + \mathbf{k}_{\text{p}m}| = 2 k_{\text{p}}$) share a common idler mode at $\mathbf{k}_{\text{i}} = 0$, such that the four corresponding signal modes are expected to be entangled.

%_______________________________________________________________________________
%-------------------------------------------------------------------------------
\subsection{Emission of light}
In the following we calculate the state of the emitted polaritons in the absence of noise or losses.
Since we consider a setup, where the lower polariton branch is resonantly excited, we can neglect all processes where polaritons from the upper polariton branch scatter into some final states.
Thus, we can assume $j_3 = j_4 = 1$ in the polariton pair interaction Hamiltonian, Eq.~\eqref{HPP}.
Inspection of the polariton dispersions in Fig.~\ref{Fig:setup} inset (a) shows that there is no energy and momentum conserving process, where two lower polaritons with incident angles below the magic angle scatter into one lower and one upper polariton.
Hence, we can neglect all contributions from the upper polariton branch and approximate the polariton-polariton interaction Hamiltonian equation~\eqref{HPP} by the parametric Hamiltonian
\begin{eqnarray}
 H_{\text{PP}}^{\text{par}} &=& \frac{1}{2} \sum_\mathbf{k} \sum_{n,m=1}^4 V_{\mathbf{k}_{\text{p}n},\mathbf{k}_{\text{p}m},\mathbf{k}-\mathbf{k}_{\text{p}n}}^{1 1 1 1} P_{1 \mathbf{k}_{\text{p}n}} P_{1 \mathbf{k}_{\text{p}m}} \nonumber\\*
 && \times p_{1 \mathbf{k}}^\dagger p_{1 \mathbf{k}_{\text{p}n}+\mathbf{k}_{\text{p}m}-\mathbf{k}}^\dagger + \text{H.\,c.} \;,
\end{eqnarray}
where $P_{1 \mathbf{k}_{\text{p}}}^2 = \langle p_{1 \mathbf{k}_{\text{p}}} \rangle^2 R_{\text{X}}^2 / A$ is a classical pump field.
For the proposed parametric scattering process with $\mathbf{k}_{\text{i}} = 0$ and $\mathbf{k}_{\text{s}1} = (0, 2k_{\text{p}})$, $\mathbf{k}_{\text{s}2} = (-2k_{\text{p}}, 0)$, $\mathbf{k}_{\text{s}3} = (0, -2k_{\text{p}})$, and $\mathbf{k}_{\text{s}4} = (2k_{\text{p}}, 0)$, the effective branch-dependent potential $V_{\mathbf{k}_{\text{p}n},\mathbf{k}_{\text{p}m},-\mathbf{k}_{\text{p}n}}^{1 1 1 1}$ for neighboring pumps can be simplified, which yields
\begin{eqnarray}
 \frac{V_{\mathbf{k}_{\text{p}n},\mathbf{k}_{\text{p}m},-\mathbf{k}_{\text{p}n}}^{1 1 1 1}}{E_{\text{b}}} &=& 12 M_{1 1 0} M_{1 1 2k_{\text{p}}} M_{1 1 \sqrt{2}k_{\text{p}}} M_{1 1 \sqrt{2}k_{\text{p}}} \nonumber\\*
 && -\frac{8 \pi}{7} p_{\text{s}} \Big( M_{2 1 0} M_{1 1 2k_{\text{p}}} M_{1 1 \sqrt{2}k_{\text{p}}} M_{1 1 \sqrt{2}k_{\text{p}}} \nonumber\\*
 && + M_{1 1 0} M_{1 1 2k_{\text{p}}} M_{1 1 \sqrt{2}k_{\text{p}}} M_{2 1 \sqrt{2}k_{\text{p}}} \Big) \;,
\end{eqnarray}
being independent of the directions of the respective wave vectors.
In obtaining this equation we took into account that the coefficients $M_{j,j',\mathbf{k}}$ depend on the modulus $k$ only.
Consequently, we may take $V_{k_{\text{p}}} = V_{\mathbf{k}_{\text{p}n},\mathbf{k}_{\text{p}m},-\mathbf{k}_{\text{p}n}}^{1 1 1 1}$ and find
\begin{eqnarray}\label{HPP_par}
 H_{\text{PP}}^{\text{par}} &=& V_{k_{\text{p}}} \big( P_{1 \mathbf{k}_{\text{p}1}} P_{1 \mathbf{k}_{\text{p}2}} p_{1 \mathbf{k}_{\text{i}}}^\dagger p_{1 \mathbf{k}_{\text{s}1}}^\dagger + P_{1 \mathbf{k}_{\text{p}2}} P_{1 \mathbf{k}_{\text{p}3}} p_{1 \mathbf{k}_{\text{i}}}^\dagger p_{1 \mathbf{k}_{\text{s}2}}^\dagger \nonumber\\*
 && + P_{1 \mathbf{k}_{\text{p}3}} P_{1 \mathbf{k}_{\text{p}4}} p_{1 \mathbf{k}_{\text{i}}}^\dagger p_{1 \mathbf{k}_{\text{s}3}}^\dagger + P_{1 \mathbf{k}_{\text{p}1}} P_{1 \mathbf{k}_{\text{p}4}} p_{1 \mathbf{k}_{\text{i}}}^\dagger p_{1 \mathbf{k}_{\text{s}4}}^\dagger \big) \nonumber\\*
 && + \text{H.\,c.}
\end{eqnarray}

We now assume coherent pump polariton fields of equal amplitude, $P_{1 \mathbf{k}_{\text{p}n}} = P_{1 \mathbf{k}_{\text{p}m}} = P_{1 k_{\text{p}}}$ for $n, m = 1, \ldots, 4$.
Then, in the limit of low excitation intensity ($V_{k_{\text{p}}} |P_{1 k_{\text{p}}}|^2 t \ll 1$)~\cite{SDSSL05}, the Hamiltonian $H_{\text{PP}}^{\text{par}}$ from Eq.~\eqref{HPP_par}---when acting on a vacuum state---generates polaritons in the state
\begin{equation}\label{psi_pol}
 | \psi_{\text{out}} \rangle = \frac{1}{2} | 1 \rangle_{\text{i}} \Big( | 1 \rangle_{\text{s}1} + | 1 \rangle_{\text{s}2} + | 1 \rangle_{\text{s}3} + | 1 \rangle_{\text{s}4} \Big) \;,
\end{equation}
where we denote with $| 1 \rangle_{\text{x}}$ the state of a polariton in channel $\text{x} = \text{i (idler)}, \text{s}1, \ldots, \text{s}4 \text{ (signal)}$.
We now take the partial trace over the idler mode $i$ to obtain the state $\rho$ of the four signal fields
\begin{equation}
 \rho = {\mathop{\text{Tr}}}_{\text{i}} | \psi_{\text{out}} \rangle \langle \psi_{\text{out}} | = | \psi \rangle \langle \psi | \;,
\end{equation}
where
\begin{equation}\label{Wstate}
 | \psi \rangle = \frac{1}{2} \big( | 1, 0, 0, 0 \rangle + | 0, 1, 0, 0 \rangle + | 0, 0, 1, 0 \rangle + | 0, 0, 0, 1 \rangle \big)
\end{equation}
is a four-partite entangled pure state, called the $W$ state.
In the following, we denote this state as a four-mode $W$ state.
In a similar way one may generate not only four-mode, but also $2^N$-mode $W$ states.

To obtain the state of the emitted light we have to couple the intracavity polariton scattering dynamics to an extracavity field and determine the parametric luminescence.
As is known from previous results~\cite{HWSOPI94,STPQS96,CSQ01} there is a correspondence between the properties of the polaritons within the cavity and the emitted photons outside the cavity.
In particular, due to energy and momentum conservation, the emitted photon has both the energy and the in-plane momentum of the corresponding polariton.
Since the coupling strength between an extracavity photon and an intracavity polariton depends on the energy and the modulus of the in-plane wave vector only, in the considered setup every (polariton) signal mode is equally coupled to a corresponding mode of the external field.
Hence, we can assume that the emitted signal fields outside the microcavity are given by the state~\eqref{psi_pol}.

%_______________________________________________________________________________
%===============================================================================
\section{Identification of multipartite entanglement}\label{Sec:IdentEnt}
In the bipartite case, several approaches for the identification and even for the quantification of entanglement exist (see, e.g., \cite{HHHH09,NC10}).
Examples are the relative entropy of entanglement~\cite{VP98} and Schmidt number witnesses for mixed states~\cite{HHH96,SV11a}.
In the $N$-partite case with $N > 2$ we must distinguish between partially and fully entangled states.
On the one hand, a pure quantum state is called partially entangled if it cannot be written as a product of states of each subsystem, i.e., it is not fully separable.
On the other hand, if the state is not even partially separable, i.\,e. we can not separate any subsystem, it is called fully entangled.
These definitions can also be extended to mixed quantum states, since they can be written as classical mixtures of pure states.

For the identification of entanglement we use the method of multipartite entanglement witnesses~\cite{SV13}.
A quantum state $\rho$ is partially entangled if and only if there exists a Hermitian operator $L$ with
\begin{equation}\label{test_part}
 \langle L \rangle = \mathop{\text{Tr}} \rho L > f_{\text{full}}(L) \;,
\end{equation}
where $f_{\text{full}}(L)$ is the maximum expectation value of $L$ for fully separable states,
\begin{equation}
 f_{\text{full}}(L) = \sup \{ \langle \psi | L | \psi \rangle : | \psi \rangle ~ \text{fully separable} \} \;.
\end{equation}
Accordingly, the state $\rho$ is fully entangled if and only if there exists a Hermitian operator $L$ with
\begin{equation}\label{test_full}
 \langle L \rangle = \mathop{\text{Tr}} \rho L > f_{\text{part}}(L) \;,
\end{equation}
where $f_{\text{part}}(L)$ is the maximum expectation value of $L$ for partially separable states,
\begin{equation}
 f_{\text{part}}(L) = \sup \{ \langle \psi | L | \psi \rangle : | \psi \rangle ~ \text{partially separable} \} \;.
\end{equation}
An entanglement witness can be constructed as $f_{\text{full/part}}(L) \mathbb{I} - L$, where $\mathbb{I}$ denotes the identity~\cite{SV09a}.

The calculation of the values of the functions $f_{\text{full}}(L)$ and $f_{\text{part}}(L)$ is based on the solution of so called separability eigenvalue (SE) equations~\cite{SV13}.
In the $N$-partite case with the combined Hilbert space $\mathcal{H} = \bigotimes_{k = 1}^N \mathcal{H}_k$ the maximum expectation value of $L$ for fully separable states can be obtained from the solution of the equations
\begin{equation}\label{SEEQfull}
 L_{\psi_1, \ldots, \psi_{k-1}, \psi_{k+1}, \ldots, \psi_N} | \psi_k \rangle = g | \psi_k \rangle
\end{equation}
for $k = 1, \ldots, N$.
Here $| \psi_k \rangle \in \mathcal{H}_k$ are the normalized eigenstates of the reduced operator
\begin{multline}
 L_{\psi_1, \ldots, \psi_{k-1}, \psi_{k+1}, \ldots, \psi_N} \\
 = \text{Tr}_{1, \ldots, k-1, k+1, \ldots, N} \Big[ \big( | \psi_1 \rangle \langle \psi_1 | \otimes \cdots \otimes | \psi_{k-1} \rangle \langle \psi_{k-1} | \\
 \otimes \mathbb{I}_k \otimes | \psi_{k+1} \rangle \langle \psi_{k+1} | \otimes \cdots \otimes | \psi_N \rangle \langle \psi_N | \big) L \Big] \;,
\end{multline}
and the corresponding eigenvalue $g$,
\begin{equation}
 g = \langle \psi_1, \ldots, \psi_N| L | \psi_1, \ldots, \psi_N \rangle,
\end{equation}
is called SE of $L$.
The value of the function $f_{\text{full}}(L)$ then is
\begin{equation}
 f_{\text{full}}(L) = \max \{ g : g ~ \text{SE of} ~ L \} \;.
\end{equation}
The value of the function $f_{\text{part}}(L)$ in general depends on the chosen decomposition of the combined Hilbert space, which can be computed by the same form of equations~\cite{SV13}.
In the $N$-partite case $\mathcal{H} = \bigotimes_{k=1}^N \mathcal{H}_k$ a separation of $\mathcal{H}$ into $K < N$ subsystems results in the SE equations
\begin{equation}\label{SEEQpart}
 L_{\psi_1, \ldots, \psi_{k-1}, \psi_{k+1}, \ldots, \psi_K} | \psi_k \rangle = g | \psi_k \rangle
\end{equation}
for $j = 1, \ldots, K$.
Note that the symbols $| \psi_k \rangle$ in Eq.~\eqref{SEEQpart} denote a state of the $k$th subsystem with $k \in [1, K]$, whereas in Eq.~\eqref{SEEQfull} it is used for a state of the $k$th mode with $k \in [1, N]$.
To obtain the value of the function $f_{\text{part}}(L)$ we have to consider all possible partial decompositions of the combined Hilbert space.
For every decomposition we calculate the maximum SE $g$.
Then, $f_{\text{part}}(L)$ is the maximum of all these values.

As an example, let us consider the four-mode $W$ state $| \psi \rangle$ from Eq.~\eqref{Wstate}.
A general pure state of the $n$th signal mode ($n = 1, \ldots, 4$) is given by $| \psi_n \rangle = (\alpha_0^n)^* | 0 \rangle + (\alpha_1^n)^* | 1 \rangle$ with $|\alpha_0^n|^2 + |\alpha_1^n|^2 = 1$.
We choose $L = \rho$ as the Hermitian test operator, such that $\langle L \rangle = \mathop{\text{Tr}} \rho L = 1$.
From symmetry reasons we have to consider one component only, say the fourth one.
The SE equation for full separability then reads $L_{\psi_1,\psi_2,\psi_3} | \psi_4 \rangle = g | \psi_4 \rangle$.
We obtain
\begin{equation}
 L_{\psi_1,\psi_2,\psi_3} = | \psi_{1,2,3} \rangle \langle \psi_{1,2,3} |
\end{equation}
with
\begin{eqnarray}
 | \psi_{1,2,3} \rangle &=& \frac{1}{2} \big( \alpha_0^1 \alpha_0^2 \alpha_1^3 + \alpha_0^1 \alpha_1^2 \alpha_0^3 + \alpha_1^1 \alpha_0^2 \alpha_0^3 \big) | 0 \rangle \nonumber\\
 && + \frac{1}{2} \alpha_0^1 \alpha_0^2 \alpha_0^3 | 1 \rangle \;.
\end{eqnarray}
Since we are interested in non-trivial solutions of the SE equation, we find $| \psi_4 \rangle = M | \psi_{1,2,3} \rangle$ with a normalization constant $M$.
The corresponding eigenvalue $g$ then is
\begin{equation}
 g = \frac{1}{4} \Big( \big| \alpha_0^1 \alpha_0^2 \alpha_1^3 + \alpha_0^1 \alpha_1^2 \alpha_0^3 + \alpha_1^1 \alpha_0^2 \alpha_0^3 \big|^2 + \big| \alpha_0^1 \alpha_0^2 \alpha_0^3 \big|^2 \Big) \;.
\end{equation}
The maximum $g$ is obtained for $\alpha_0^1 = \alpha_0^2 = \alpha_0^3 = \sqrt{3} / 2$ and we find
\begin{equation}
 f_{\text{full}}(L) = \frac{27}{64} \;,
\end{equation}
which obviously is smaller than one, such that the four-mode $W$ state is shown to be partially entangled.

As mentioned above, the SE equations for partial separability depend on the chosen separation.
In a first step, we consider the fourth mode as separated.
The solution of the corresponding SE equation then yields the maximum expectation value $f_{\text{part}}^{123:4}(L) = 3 / 4$, where the super-index 123:4 indicates the chosen decomposition.
For symmetry reasons permutations of this separation will result in the same value.
For the remaining separations we find $f_{\text{part}}^{12:3:4}(L) = f_{\text{part}}^{12:34}(L) = 1 / 2$.
Hence,
\begin{equation}
 f_{\text{part}}(L) = \frac{3}{4} \;,
\end{equation}
such that the four-mode $W$ state $| \psi \rangle$ from Eq.~\eqref{Wstate} is not only partially but also fully entangled.

%_______________________________________________________________________________
%===============================================================================
\section{Entanglement in the presence of losses}\label{Sec:EntProp}
In this section we study the propagation of entangled light through media which can be described by realistic loss models, cf.~\cite{VW06,SV10atm,VSV12}.
This may include losses during the outcoupling of the field from the cavity~\cite{KVW91}, and the subsequent propagation through lossy media.
Of special importance are turbulent media since they describe the typical propagation of light in the atmosphere~\cite{SV09atm}.
In particular, we perform an entanglement test where the witness is based on a general pure $W$ state.
This allows us to study the effects of the lossy channel on the entanglement within the state of the signal fields.

%_______________________________________________________________________________
%-------------------------------------------------------------------------------
\subsection{Mixing with vacuum}
We consider the case of a four-mode radiation field with up to one photon per mode.
In a random loss model the initial pure state mixes with some vacuum contributions.
A replacement scheme of this atmospheric propagation is a chain of beam splitters, which transmits a part of the incident light and scatters the remaining radiation.
The scattered part of the light is given by the reflectivity $\sqrt{1 - \eta}$ of the beam splitter and, in general, depends on the wave vector $\mathbf{k}$ of the propagating light, i.e., the quantum efficiency is~$\eta = \eta(\mathbf{k})$.

Mathematically, this process can be described by replacing the polariton creation operators $p_{1 \mathbf{k}_{\text{s}n}}$ in the Hamiltonian equation~\eqref{HPP_par} by a loss model of the output light $\sqrt{\eta_n} p_{\mathbf{k}_{\text{s}n}}^\dagger + \sqrt{1 - \eta_n} b_n^\dagger$, where $\eta_n = \eta(\mathbf{k}_{\text{s}n})$ and $b_n^\dagger$ creates a polariton in bath $n$.
Note that the elaboration for polaritons instead of photons is justified through the equivalence of their respective momenta~\cite{STPQS96,SDSSL05}, as we already mentioned in Sec.~\ref{Sec:EntGen}.
The state of the emitted polaritons is then obtained by applying the resulting Hamiltonian onto the polariton vacuum.
To obtain the state of the signal fields only, we take the partial trace over the idler mode and the bath degrees of freedom.
The resulting state reads
\begin{equation}\label{rho_mix}
 \rho_{\text{mix}} = | \psi_{\text{mix}} \rangle \langle \psi_{\text{mix}} | + \frac{1}{4} (4 - \sum_{n=1}^4 \eta_n) | 0, 0, 0, 0 \rangle \langle 0, 0, 0, 0 | \;,
\end{equation}
where
\begin{eqnarray}
 | \psi_{\text{mix}} \rangle &=& \frac{1}{2} \big( \sqrt{\eta_1} | 1, 0, 0, 0 \rangle + \sqrt{\eta_2} | 0, 1, 0, 0 \rangle \nonumber\\*
 && + \sqrt{\eta_3} | 0, 0, 1, 0 \rangle + \sqrt{\eta_4} | 0, 0, 0, 1 \rangle \big)
\end{eqnarray}
is a generalized four-mode $W$ state.
Let us also note that the turbulence model of losses is given by a probability distribution $\mathcal P(\eta_1,\ldots,\eta_4)$ of the quantum efficiencies~\cite{SV09atm}.
In the considered approximation this yields a replacement of the values $\eta_n$ with the corresponding mean values.

%_______________________________________________________________________________
%-------------------------------------------------------------------------------
\subsection{General $W$-state witness}\label{Sec:WWitness}
The mixed four-mode state $\rho_{\text{mix}}$ from Eq.~\eqref{rho_mix} can be written as the trace of a pure five-mode state over the fifth mode, i.e.,
\begin{equation}
 \rho_{\text{mix}} = \text{Tr}_5 | W_5 \rangle \langle W_5 | \;.
\end{equation}
Here,
\begin{eqnarray}\label{5modeW}
 | W_5 \rangle &=& \frac{\sqrt{\eta_1}}{2} | 1, 0, 0, 0, 0 \rangle + \frac{\sqrt{\eta_2}}{2} | 0, 1, 0, 0, 0 \rangle \nonumber\\
 && + \frac{\sqrt{\eta_3}}{2} | 0, 0, 1, 0, 0 \rangle + \frac{\sqrt{\eta_4}}{2} | 0, 0, 0, 1, 0 \rangle \nonumber\\
 && + \frac{1}{2} \sqrt{4 - \eta_1 - \eta_2 - \eta_3 - \eta_4} | 0, 0, 0, 0, 1 \rangle
\end{eqnarray}
is a generalized five-mode $W$ state.
In order to detect entanglement within the state $\rho_{\text{mix}}$, i.e., in order to calculate the right-hand sides of the conditions~\eqref{test_part} and~\eqref{test_full}, it is therefore sufficient to consider a test operator based on a pure $W$ state only.
This property is known as the theorem of cascaded structures~\cite{SV13}.
Since the right-hand sides of the entanglement conditions~\eqref{test_part} and~\eqref{test_full} are independent of the considered state, the results can be used to detect entanglement for any arbitrary state.
Thus, we here consider the general test operator $L = | W_N \rangle \langle W_N |$ based on the generalized $N$-mode $W$ state
\begin{equation}
 | W_N \rangle = \sum_{i=1}^N \lambda_i | \underbrace{0, \ldots, 0}_{i-1}, 1, \underbrace{0, \ldots, 0}_{N-i} \rangle \;,
\end{equation}
where the $\lambda_i$ for $i = 1, \ldots, N$ are the weights of the respective modes.

%_______________________________________________________________________________
%...............................................................................
\subsubsection{Test for partial entanglement}
As mentioned before, the value of the function $f_{\text{full}}(L)$ is obtained from the solution of the corresponding separability eigenvalue equations~\eqref{SEEQfull}.
For the state $| \psi_n \rangle$ of the $n$th subsystem we choose the parametrization $| \psi_n \rangle = (\alpha_0^n)^* | 0 \rangle + (\alpha_1^n)^* | 1 \rangle$ with the normalization $|\alpha_0^n|^2 + |\alpha_1^n|^2 = 1$.
Explicitly, we get $L_{\psi_1,\ldots,\psi_{k-1},\psi_{k+1},\ldots,\psi_N} = | \psi_{1,\ldots,k-1,k+1,\ldots,N} \rangle \langle \psi_{1,\ldots,k-1,k+1,\ldots,N} |$ with
\begin{equation}
 | \psi_{1,\ldots,k-1,k+1,\ldots,N} \rangle = \sum_{\substack{i=1 \\ i \neq k}}^N \lambda_i \alpha_1^i \prod_{\substack{j=1 \\ j \neq i,k}}^N \alpha_0^j | 0 \rangle + \lambda_k \prod_{\substack{j=1 \\ j \neq k}}^N \alpha_0^j | 1 \rangle \;,
\end{equation}
and
\begin{equation}
 g = \Bigg| \sum_{i=1}^N \lambda_i \alpha_1^i \prod_{\substack{j=1 \\ j \neq i}}^N \alpha_0^j \Bigg|^2 \;.
\end{equation}
This expression has to be maximized over all $\alpha_0^n$ and $\alpha_1^n$.
We may decompose $\alpha_x^n = r_x^n \rme^{\rmi \varphi_x^n}$ and $\lambda_n = |\lambda_n| \rme^{\rmi \theta_n}$ for $x = 0,1$ and $n = 1, \ldots, N$ in polar coordinates, such that $r_1^n = \sqrt{1 - (r_0^n)^2}$.
The definition $r_n = r_0^n$ and the maximization over all $\varphi_x^n$ and $\theta_n$ then leads to the equation
\begin{equation}\label{g_FullSep}
 g = \Bigg( \sum_{i=1}^N |\lambda_i| \sqrt{1 - r_i^2} \prod_{\substack{j=1 \\ j \neq i}}^N r_j \Bigg)^2 \;.
\end{equation}
We now have to maximize over $r_n \in [0, 1]$ for $n = 1, \ldots, N$.
At the borders $r_n = 0, 1$ the function in Eq.~\eqref{g_FullSep} assumes the solutions
\begin{equation}\label{obv_sol}
 g = 0, |\lambda_1|^2, \ldots, |\lambda_N|^2 \;.
\end{equation}
If $g$ has a local maximum for at least one $r_n \in (0,1)$, the partial derivatives $\partial g / \partial r_n$ vanish at this point.
This requirement leads to the $N$ equations
\begin{equation}\label{x_FullSep}
 |\lambda_n| = x_n \sum_{\substack{i=1 \\ i \neq n}}^N |\lambda_i| x_i
\end{equation}
for $n = 1, \ldots, N$, where we introduced the new variables $x_n = \sqrt{1 - r_n^2} / r_n \in (0, \infty)$.
To obtain the global maximum of $g$ we have to compare the solutions~\eqref{obv_sol} with the local extrema determined from the solution of Eqs.~\eqref{x_FullSep}.

For general choices of the weights $\lambda_n$ for $n = 1, \ldots, N$ Eqs.~\eqref{x_FullSep} have to be solved numerically.
Analytical results can be obtained for an equal-weighted $W$ state with $|\lambda_n| = 1 / \sqrt{N}$ for all $n = 1, \ldots, N$.
Then, the solution of Eqs.~\eqref{x_FullSep} reads
\begin{equation}
 x_1 = \ldots = x_N = \frac{1}{\sqrt{N - 1}} \;,
\end{equation}
such that
\begin{equation}
 g_{\text{max}} = f_{\text{full}}(L) = \left( \frac{N - 1}{N} \right)^{N-1} \;.
\end{equation}
A more general but also analytically solvable situation arises, if we assume that all but one weights are equal, i.e., $|\lambda_1| = \cdots = |\lambda_{N-1}| = \lambda$ and $|\lambda_N| = \lambda'$.
Note that this situation corresponds to the choice of equal reflectivities $\eta_1 = \eta_2 = \eta_3 = \eta_4$ within the five-mode $W$ state from Eq.~\eqref{5modeW}.
After some algebra, we get in the general $N$-mode case
\begin{equation}\label{gMax_oneDiff}
 g_{\text{max}} = (N - 1) \lambda^2 \left( \frac{(N - 1) (N - 2) \lambda^2}{(N - 1)^2 \lambda^2 - (\lambda')^2} \right)^{N-2} \;,
\end{equation}
which is valid for $\lambda' / \lambda < \sqrt{N - 1}$.
If $\lambda' / \lambda \geq \sqrt{N - 1}$ Eqs.~\eqref{x_FullSep} have no solution, such that $g_{\text{max}} = \max_{n=1}^N |\lambda_n|^2$ taking the solutions~\eqref{obv_sol} into account.

%_______________________________________________________________________________
%...............................................................................
\subsubsection{Test for full entanglement}
To obtain the value of the function $f_{\text{part}}(L)$ we have to consider all possible separations of the Hilbert space.
In a first step we study a general bipartite decomposition of the state $| W_N \rangle$.
In particular, we consider the $n$ subsystems with indices $k_1 < \cdots < k_n$ as one party (system $A$) and the other $N-n$ subsystems with indices $k_{n+1} < \cdots < k_N$ as the second party (system $B$).
We then may write
\begin{equation}
 | W_N \rangle = \sum_{i=1}^n \lambda_{k_i} | 2^i, 0 \rangle + \sum_{i=n+1}^N \lambda_{k_i} | 0, 2^{i-n} \rangle \;,
\end{equation}
where we introduced the abbreviations
\begin{equation}\label{systemA}
 | 2^i, 0 \rangle = | \underbrace{0,\ldots,0}_{k_i-1}, 1, \underbrace{0,\ldots,0}_{N-k_i} \rangle \;,
\end{equation}
and
\begin{equation}\label{systemB}
 | 0, 2^{i-n} \rangle = | \underbrace{0,\ldots,0}_{k_i-1}, 1, \underbrace{0,\ldots,0}_{N-k_i} \rangle
\end{equation}
in the form of two binary numbers for the states of the two parties.
Note that, although the right-hand sides of Eqs.~\eqref{systemA} and~\eqref{systemB} look equal, they belong to different states.
In Eq.~\eqref{systemA} the mode $k_i$ corresponds to a subsystem of system $A$ ($i \in [1, n]$), whereas in Eq.~\eqref{systemB} the $k_i$ belongs to a subsystem of system $B$ ($i \in [n+1, N]$).

We trace out the system $A$ and obtain
\begin{equation}
 \text{Tr}_A L = \sum_{i,j=n+1}^N \lambda_{k_i}^{} \lambda_{k_j}^* | 2^{i-n} \rangle \langle 2^{j-n} | + \sum_{i=1}^n |\lambda_{k_i}|^2 | 0 \rangle \langle 0 | \;.
\end{equation}
Since this expression already is the spectral decomposition, i.e., it is diagonal in the two states
\begin{equation}
 | 0 \rangle \;, \qquad
 \bigg( \sum_{i=n+1}^N |\lambda_{k_i}|^2 \bigg)^{-1/2} \sum_{i=n+1}^N \lambda_{k_i} | 2^{i-n} \rangle \;,
\end{equation}
we get two separability eigenvalues for the considered bipartition, such that
\begin{equation}\label{g_bipartite}
 g_{\text{max}} = \max \left\{ \sum_{i=1}^n |\lambda_{k_i}|^2 \;, ~ \sum_{i=n+1}^N |\lambda_{k_i}|^2 \right\} \;.
\end{equation}
Note that the method of tracing out a system and reading of the separability eigenvalues from the result is valid only in the bipartite case ($N = 2$) because in this situation the solutions~\eqref{obv_sol} are the only ones.

The maximum of the values~\eqref{g_bipartite} is obtained for $n = 1$, if we choose the mode with the smallest weight as system $A$, or for $n = N - 1$, if we choose the $N - 1$ modes with the largest weights as system $A$.
The resulting eigenvalue is then the sum of the $N - 1$ largest $|\lambda_i|^2$.
In particular, in the case of equal weights we have $g_{\text{max}} = (N - 1) / N$, and in the case of all but one equal weights we have $g_{\text{max}} = (N - 1) \lambda^2$ if $\lambda > \lambda'$ and $g_{\text{max}} = (N - 2) \lambda^2 + (\lambda')^2$ if $\lambda < \lambda'$.

We now consider the general decomposition of the combined Hilbert space into $K$ subsystems.
We may write
\begin{equation}
 | W_N \rangle = \sum_{n=1}^K \sum_{m=1}^{N_n} \lambda_m^{(n)} | 0, \ldots, 0, 2^m, 0, \ldots, 0 \rangle \;,
\end{equation}
where $N_n$ is the number of modes that are combined into subsystem $n$.
Accordingly, $\lambda_m^{(n)}$ denotes the weight of the $m$th mode within subsystem $n$.
The state $| 0, \ldots, 0, 2^m, 0, \ldots, 0 \rangle$ is a product of states of the $n$ subsystems, where $0$ denotes the respective vacuum state and $2^m$ denotes the state of one photon in mode $m$.
To shorten the expressions we introduce
\begin{equation}\label{a_Ksep}
 | a_n \rangle = \sum_{m=1}^{N_n} \lambda_m^{(n)} | 2^m \rangle \;,
\end{equation}
being the (unnormalized) state of the $n$th subsystem, such that
\begin{equation}
 | W_N \rangle = \sum_{n=1}^K | 0, \ldots, 0, a_n, 0, \ldots, 0 \rangle \;.
\end{equation}

Similar to the case of partial entanglement we have to solve the separability eigenvalue equations~\eqref{SEEQpart} and obtain $L_{\psi_1,\ldots,\psi_{n-1},\psi_{n+1},\ldots,\psi_K} = | \psi_{1,\ldots,n-1,n+1,\ldots,K} \rangle \langle \psi_{1,\ldots,n-1,n+1,\ldots,K} |$ with
\begin{eqnarray}
 | \psi_{1,\ldots,n-1,n+1,\ldots,K} \rangle &=& \Bigg( \sum_{\substack{i=1 \\ i \neq n}}^K \langle \psi_i | a_i \rangle \prod_{\substack{j=1 \\ j \neq i,n}}^K \langle \psi_j | 0 \rangle \Bigg) | 0 \rangle \nonumber\\
 && + \Bigg( \prod_{\substack{j=1 \\ j \neq n}}^K \langle \psi_j | 0 \rangle \Bigg) | a_n \rangle \;.
\end{eqnarray}
Hence, we can use
\begin{equation}
 | \psi_n \rangle = (\alpha_0^n)^* | 0 \rangle + \frac{(\alpha_1^n)^*}{\sqrt{M_n}} | a_n \rangle
\end{equation}
with $|\alpha_0^n|^2 + |\alpha_1^n|^2 = 1$ and $M_n = \langle a_n | a_n \rangle$ as parametrization for the state of the $n$th subsystem.
The separability eigenvalue then follows as
\begin{equation}
 g = \Bigg| \sum_{i=1}^K \sqrt{M_i} \alpha_1^i \prod_{\substack{j=1 \\ j \neq i}}^K \alpha_0^j \Bigg|^2 \;.
\end{equation}
Again, we can decompose $\alpha_x^n = r_x^n \rme^{\rmi \varphi_x^n}$ for $x = 0, 1$ and $n = 1, \ldots, K$ in polar coordinates such that $r_1^n = \sqrt{1 - (r_0^n)^2}$, define $r_n = r_0^n$, and obtain after maximization over the phases $\varphi_x^n$ the relation
\begin{equation}\label{g_Ksep}
 g = \Bigg( \sum_{i=1}^K \sqrt{M_i} \sqrt{1 - r_i^2} \prod_{\substack{j=1 \\ j \neq i}}^K r_j \Bigg)^2 \;.
\end{equation}
The solutions $g = M_n$ for $n = 1, \ldots, K$ are obtained for $r_n = 0$ and $r_i = 1$ for $i = 1, \ldots, K$ with $i \neq n$.
For all other solutions we can compare Eq.~\eqref{g_Ksep} with Eq.~\eqref{g_FullSep} to see that the structure of both expressions is the same.
It follows that the maximum separability eigenvalue in the case of full entanglement can be obtained from the general solution to Eq.~\eqref{g_FullSep} for partial entanglement if we replace the number of modes $N$ by the number of subsystems $K$ and the weights $|\lambda_i|$ by the values $\sqrt{M_i}$.

%_______________________________________________________________________________
%-------------------------------------------------------------------------------
\subsection{Analytical and numerical results}
The general results from the last section allow us to identify the parameter range of the efficiencies $\eta_n$, for which the mixed signal state $\rho_{\text{mix}}$ from Eq.~\eqref{rho_mix} is partially and fully entangled.
Choosing $L = \rho_{\text{mix}}$ we obtain for the left-hand side of all tests
\begin{equation}\label{TrRhoL_res}
 \mathop{\text{Tr}} \rho_{\text{mix}} L = 1 - \frac{1}{2} \sum_{n=1}^4 \eta_n + \frac{1}{8} \bigg( \sum_{n=1}^4 \eta_n \bigg)^2 \;.
\end{equation}

Analytical results can be obtained for equal reflectivities, i.e., $\eta_n = \eta$ for all $n = 1, \ldots, 4$.
Because the moduli of the signal wave vectors $\mathbf{k}_{\text{s}n}$ are equal in the considered scenario this assumption corresponds to a situation, where the reflectivities are isotropic.
Then, we can use the result from Eq.~\eqref{gMax_oneDiff} yielding
\begin{equation}\label{fFull_res}
 f_{\text{full}}(L) = \begin{cases} 1 - \eta & 0 \leq \eta \leq 1/2 \\ 27 \eta^4 (5 \eta - 1)^{-3} & 1/2 < \eta \leq 1 \end{cases} \;.
\end{equation}
To determine the value of the function $f_{\text{part}}(L)$ we have to solve Eq.~\eqref{g_Ksep} for all relevant decompositions of the combined Hilbert space.
Since we use the operator $L = | W_5 \rangle \langle W_5 |$ based on the pure five-mode $W$ state from Eq.~\eqref{5modeW} instead of the mixed four-mode state $\rho_{\text{mix}}$ from Eq.~\eqref{rho_mix}, the fifth mode should always be considered as a separated party.
For the remaining four modes we have to allow for all possible decompositions.
As a result, the maximum SE is obtained if we consider the fourth and the fifth mode as separated, such that the states $| a_n \rangle$ from Eq.~\eqref{a_Ksep} are given by $| a_1 \rangle = (\sqrt{\eta} / 2) (| 1, 0, 0 \rangle + | 0, 1, 0 \rangle + | 0, 0, 1 \rangle)$, $| a_2 \rangle = (\sqrt{\eta} / 2) | 1 \rangle$, and $| a_3 \rangle = \sqrt{1 - \eta} | 1 \rangle$.
It follows that $M_1 = 3 \eta / 4$, $M_2 = \eta / 4$, and $M_3 = 1 - \eta$.
After maximization of Eq.~\eqref{g_Ksep} for these values we obtain
\begin{equation}\label{fPart_res}
 f_{\text{part}}(L) = \begin{cases} 1 - \eta & 0 \leq \eta \leq 1/2 \\ \dfrac{3 \eta^2 (\eta - 1)}{13 \eta^2 - 16 \eta + 4} & 1/2 < \eta < 2/3 \\ 3 \eta / 4 & 2/3 \leq \eta \leq 1 \end{cases} \;.
\end{equation}
Note that according to our assumption of equal reflectivities the same result is obtained if one considers the first, second or third together with the fifth mode as separated.

\begin{figure}
 \includegraphics[width=0.4\textwidth]{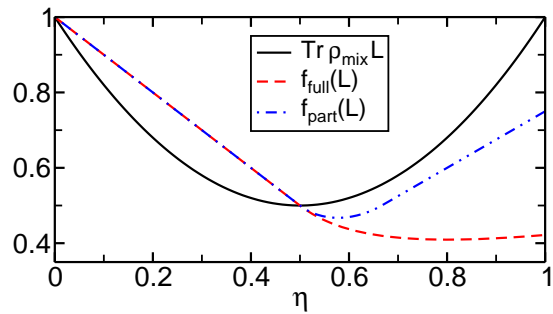}
 \caption{\label{Fig:MixVac_Lrho}
  (Color online) Left- and right-hand sides of Eqs.~\eqref{test_part} and~\eqref{test_full} as functions of $\eta = \eta_n$ for $n = 1, \ldots, 4$.
  The state $\rho_{\text{mix}}$ is partially entangled in regions where $\mathop{\text{Tr}} \rho_{\text{mix}} L > f_{\text{full}}(L)$ and fully entangled if $\mathop{\text{Tr}} \rho_{\text{mix}} L > f_{\text{part}}(L)$ with respect to the choice $L = \rho_{\text{mix}}$.
 }
\end{figure}

In Fig.~\ref{Fig:MixVac_Lrho} we show the results $f_{\text{full}}(L)$ and $f
_{\text{part}}(L)$ from Eqs.~\eqref{fFull_res} and~\eqref{fPart_res} together with $\mathop{\text{Tr}} \rho_{\text{mix}} L$ from Eq.~\eqref{TrRhoL_res} as functions of $\eta$.
We see, that we can detect entanglement for $\eta > 1/2$.
The witness based on $L = \rho_{\text{mix}}$ does not distinguish between partial and full entanglement.

\begin{figure}
 \includegraphics[scale=0.34]{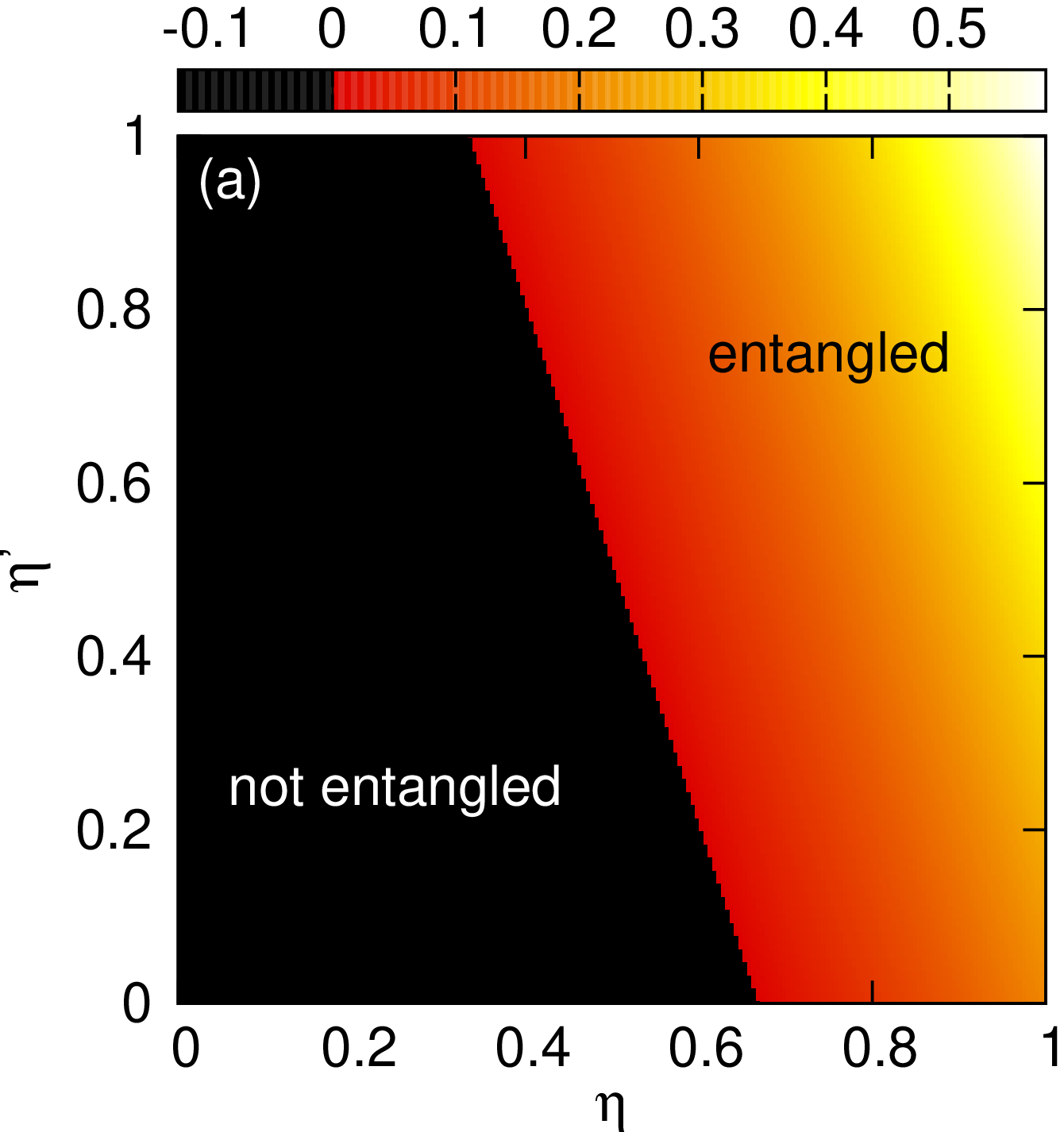}~~\includegraphics[scale=0.34]{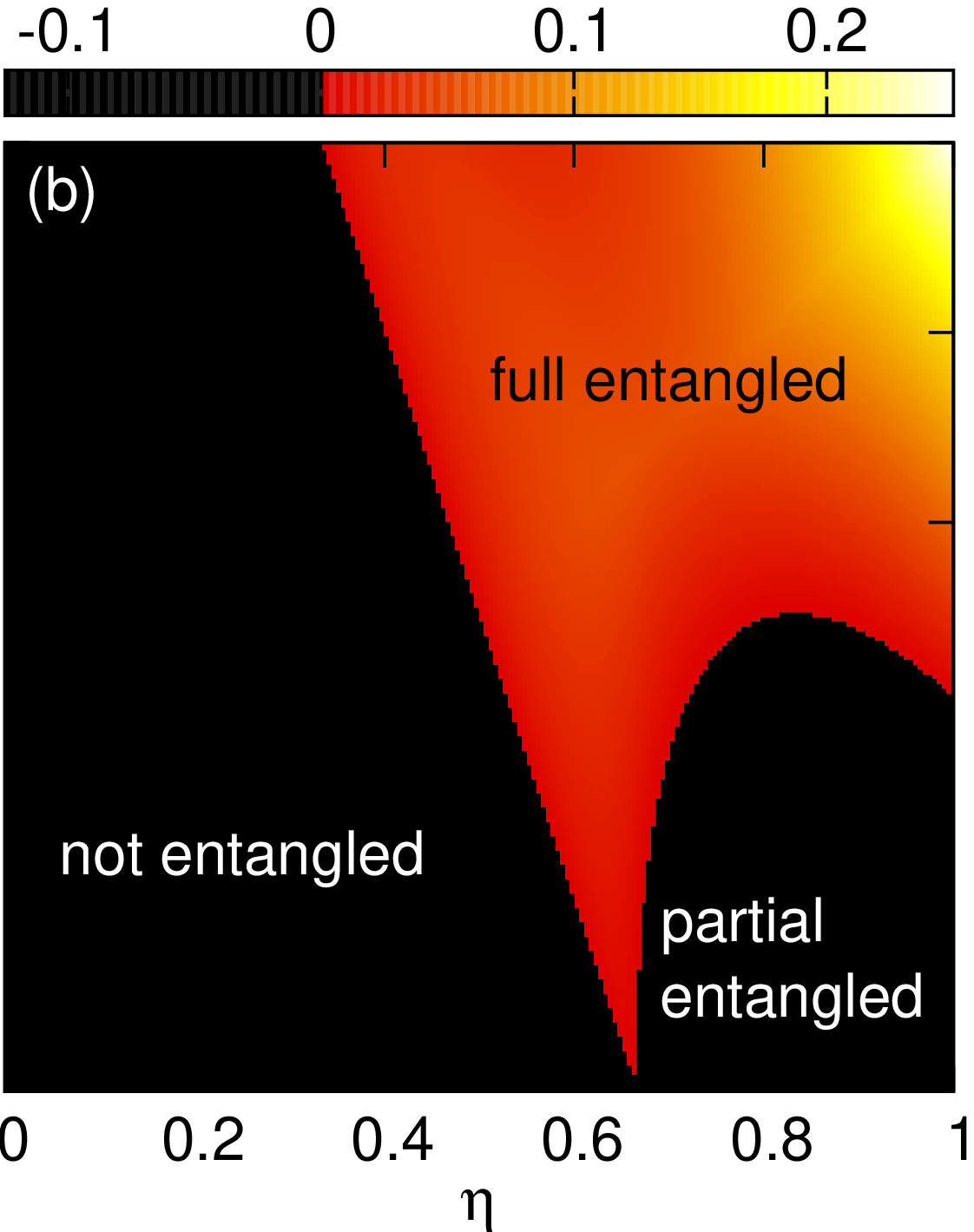}
 \caption{\label{Fig:TrRhoL_f}
  (Color online) Conturplots of (a) $\mathop{\text{Tr}} \rho_{\text{mix}} L - f_{\text{full}}(L)$ and (b) $\mathop{\text{Tr}} \rho_{\text{mix}} L - f_{\text{part}}(L)$ as functions of $\eta = \eta_1 = \eta_2 = \eta_3$ and $\eta' = \eta_4$.
  The state $\rho_{\text{mix}}$ is partially or fully entangled in regions where the respective function takes positive values.
 }
\end{figure}

Numerically, we can also study the case of unequal reflectivities.
As an example, we choose $\eta = \eta_n$ for $n = 1, 2, 3$ and $\eta' = \eta_4$.
This choice corresponds to a non isotropic situation, where the dependency of the reflectivity on the direction $\mathbf{k}_{\text{s}4}$ differs from that of the other directions.
In Figs.~\ref{Fig:TrRhoL_f}(a) and~\ref{Fig:TrRhoL_f}(b) we show the numerical results for the functions $\mathop{\text{Tr}} \rho_{\text{mix}} L - f_{\text{full}}(L)$ and $\mathop{\text{Tr}} \rho_{\text{mix}} L - f_{\text{part}}(L)$, respectively.
We see, that these functions take positive values in some regions indicating that the state $\rho_{\text{mix}}$ is partially or fully entangled.
From the test for partial entanglement---summarized in the left panel (a) of Fig.~\ref{Fig:TrRhoL_f}---we conclude that the state $\rho_{\text{mix}}$ is not entangled within the black region and contains some entangled modes within the colored region.
Panel (b) of Fig.~\ref{Fig:TrRhoL_f} shows the corresponding results for the test for full entanglement.
This panel shows an additional black region at the lower right corner, where the state $\rho_{\text{mix}}$ is not fully entangled.
Together with the results from panel (a) of Fig.~\ref{Fig:TrRhoL_f} we conclude that the state $\rho_{\text{mix}}$ is partially entangled in this additional black region and fully entangled within the remaining colored region.

%_______________________________________________________________________________
%===============================================================================
\section{Conclusions}\label{Sec:Concl}
We have studied the generation and the characterization of multipartite entanglement of light that is emitted from a planar semiconductor microcavity.
Here, a monochromatic pumping of the lower polariton branch with four pumps arranged on a cone with opening angle below twice the magic angle leads to the emission of light in a four-mode $W$ state.
From an experimental point of view the most critical points in building the proposed setup are the realization of the pump geometry, the suppression of stray light, and the detection of the entangled output fields.
Using linear optics, such as beam splitters and mirrors, our scheme requires careful adjustment of all optical paths.
Recently, the authors of Ref.~\cite{MPSVWSLHF12} proposed a multi-dimensional laser spectroscopy setup using spatial light modulators.
Adapting this concept, it should also be possible to realize the required pump geometry.
The stray light can be suppressed by means of spatial filtering~\cite{GWUUJMH13}, such that the setup under study can be experimentally realized~\cite{privStolz}.
In the context of the detection of entanglement, the advantage of our setup is that all pumps share the same frequency.
In addition, all signal fields share the same frequency as well, which is different from the pump frequency however.
This makes it possible to perform interference experiments or balanced homodyne detection of the four signal fields~\cite{DGSHPES11}.

In our theoretical study, the identification of the multipartite entanglement of the light emitted from the microcavity is done by using the method of multipartite entanglement witnesses.
We provide the solution for an entanglement test with a witness based on a general $N$-mode $W$ state.
Using this solution, we characterize the light propagation through lossy channels regarding its entanglement properties.
We showed that we can guarantee partial and full entanglement for certain ranges of loss.
In our theoretical description the boundaries between these regions are sharp.
In an experimental realization the distance between the left-hand side, $\mathop{\text{Tr}} \rho L$, and the right-hand side, $f_{\text{full/part}}(L)$, of the entanglement condition determines the maximum allowed fluctuations for a successful entanglement test.

From our results we can conclude, that in the case of a pure state the optimal entanglement witness is given by the state itself.
Due to the theorem of cascaded structures~\cite{SV13}, we may reduce the optimal test for mixed states to a pure test with one additional degree of freedom.
This allows us to verify the entanglement of mixed states as well.

In particular, we deduce general criteria to decide whether an arbitrary $N$-mode state $\rho$ is partially or fully entangled.
For this purpose we constructed an appropriate test operator based on the $N$-mode $W$ state itself.
For every bipartite decomposition of the combined Hilbert space the corresponding boundary of the entanglement condition is readily calculated.
Thus, we can perform a test for entanglement for every bipartite decomposition of the considered state.
The full classification of the state $\rho$ is the following.
First, if there is no bipartite decomposition for which entanglement has been verified, then the state $\rho$ under study is not entangled.
Second, if there is at least one decomposition with a successful test, the state is at least partially entangled.
Third, if any test is positive for any bipartite decomposition, the state is fully entangled.
In the second case we can even identify which modes are entangled and which separate from all others.
For this task we have to gradually repeat the bipartite tests within the two subsystems that are not entangled.

%_______________________________________________________________________________
%===============================================================================
\begin{acknowledgments}
We thank H.~Stolz for valuable discussions.
This work was supported by the Deutsche Forschungsgemeinschaft through SFB 652 by projects B5 and B12.
\end{acknowledgments}

\bibliography{ref}

\end{document}